\documentclass[prd,aps]{revtex4}

\usepackage{graphicx,subfigure}

\newcommand{\be}{\begin{equation}}
\newcommand{\ee}{\end{equation}}
\newcommand{\bea}{\begin{eqnarray}}
\newcommand{\eea}{\end{eqnarray}}
\newcommand{\bd}{\begin{displaymath}}
\newcommand{\ed}{\end{displaymath}}

\newcommand{\lx}{ \ln ( 1+ (1-q)x )}
\newcommand{\ly}{ \ln ( 1+ (1-q)y )}

\newcommand{\q}{ 1+ (1-q)t }
\newcommand{\xa}{(x^{-1})_q}

\newcommand{\lc}{ \{ \ln (2-q) \} }

\newcommand{\la}{ \ln_q x }

\newcommand{\qq}{( 1+ (1-q)t )^{\frac{1}{1-q}}}
\newcommand{\qi}{ \frac{1}{1-q} }

\newcommand{\x}{ t^{1-q}  }

\newcommand{\op}{ \oplus  }
\newcommand{\om}{ \ominus  }
\newcommand{\ot}{ \otimes  }
\newcommand{\ott}{ \tilde{\otimes}  }

\newcommand{\od}{ \oslash }

\newcommand{\lb }{ \left( }
\newcommand{\rb }{ \right ) }

\begin{document}


\title{
New exponential, logarithm and q-probability in the non-extensive statistical physics }

\author{ Won Sang Chung }
\email{mimip4444@hanmail.net}

\affiliation{
Department of Physics and Research Institute of Natural Science, College of Natural Science, Gyeongsang National University, Jinju 660-701, Korea
}

\date{\today}

\begin{abstract}
In this paper, a new exponential and logarithm related to the non-extensive statistical physics is proposed by using the q-sum and q-product which satisfy the distributivity. And we discuss the q-mapping from an ordinary probability to q-probability. The q-entropy defined by the idea of q-probability is shown to be q-additive.

\end{abstract}

\maketitle

\section{Introduction}

Boltzman-Gibbs statistical mechanics shows how fast microscopic physics with short-range interaction has as effect on much larger space-time scale. The Boltzman-Gibbs entropy is given by
\be
S_{BG} = - k \sum_{i=1}^W p_i \ln p_i = k \sum_{i=1}^W   p_i\ln \frac{1}{p_i}
\ee
where $k$ is a Boltzman constant, $W$ is a total number of microscopic possibilities of the system and $p_i$ is a probability of a given microstate among $W$ different ones satisfying $ \sum_{i=1}^W p_i = 1 $. When $ p_1 = \frac{1}{W} $, we have $S_{BG} = k \ln W $.

Boltzman-Gibbs theory is not adequate for various complex, natural, artificial and social system. For instance, this theory does not explain the case that a zero maximal Lyapunov exponent appears. Typically, such situations are governed by power-laws instead of exponential distributions. In order to deal with such systems, the non-extensive statistical mechanics is proposed by C.Tsallis [1,2]. The non-extensive entropy is defined by
\be
 S_q = k ( \Sigma_{i}^W p_i^q - 1 ) / ( 1-q )
\ee
The non-extensive entropy has attracted much interest among the physicist, chemist and mathematicians who study the thermodynamics of complex system [3]. When the deformation parameter $q$ goes to 1, Tsallis entropy (2) reduces to the ordinary one (1). The non-extensive statistical mechanics has been treated along three lines:

\vspace{0.5cm}

1. Mathematical development [4, 5, 6]

2. Observation of experimental behavior [7]

3. Theoretical physics ( or chemistry) development [8]

\vspace{0.5cm}

The basis of the non-extensive statistical mechanics is q -deformed exponential and logarithmic function which is different from those of Jackson's [9]. The q-deformed exponential and q-logarithm of non-extensive statistical mechanics is defined by [10]
\be
\la = \frac{ \x -1 } {1-q}, ~~~ ( t >0 )
\ee
\be
e_q (t)  =  \qq , ~~~ (x, q \in R )
\ee
where $ \q >0 $.

From the definition of q-exponential and q-logarithm, q-sum, q-difference, q-product and q-ratio are defined by [ 5, 6]
\bea
x \op y &=& x + y + (1-q) xy \cr
x \om y &=& \frac{ x-y}{1+(1-q)y} \cr
x \ot y &=& [  x^{1-q}+ y^{1-q} -1 ]^{\qi} \cr
x \od y &=& [  x^{1-q} - y^{1-q} +1 ]^{\qi}
\eea
It can be easily checked that the operation $\op$ and $ \ot $ satisfy commutativity and associativity. For the operator $ \op$, the identity additive is $0$, while for the operator $\ot$ the identity multiplicative is $1$. Indeed, there exist an analogy between this algebraic system and the role of hyperbolic space in metric topology [11].  Two distinct mathematical tools appears in the study of physical phenomena in the complex media which is characterized by singularities in a compact space [12].

For the new algebraic operation, q-exponential and q-logarithm have the following properties:

\bea
\ln_q ( x y ) &=& \ln_q x \op \ln_q y ~~~~~ e_q (x) e_q (y) = e_q ( x \op y ) \cr
\ln_q ( x \ot y ) &=& \ln_q x + \ln_q y ~~~~~ e_q (x)\ot  e_q (y) = e_q ( x + y ) \cr
\ln_q ( x / y ) &=& \ln_q x \om \ln_q y ~~~~~ e_q (x)/  e_q (y) = e_q ( x \om y ) \cr
\ln_q ( x \od y ) &=& \ln_q x - \ln_q y ~~~~~ e_q (x)\od  e_q (y) = e_q ( x - y )
\eea

From the associativity of $\op$ and $\ot$, we have the following formula :
\be
\underbrace{t \op t \op t\op \cdots \op t }_{ n~times}= \qi \{ [\q]^n -1 \}
\ee
\be
t^{\ot^n } = \underbrace{t \ot t \ot t\ot \cdots \ot t }_{ n~times}= [n t^{1-q} - (n-1) ]^{\qi}
\ee

\section {New q-calculus}

In this section, we discuss the new algebraic operation related to the non-extensive statistical physics. The q-sum, q-difference, q-product and q-ratio defined in the eq.(5) does not obey distributivity. To resolve this problem, the new multiplication is introduced [13].

Inserting $t=1$ in eq.(7), we have
\be
\underbrace{1 \op 1 \op 1\op \cdots \op 1 }_{ n~times}= \qi \{ (2-q)^n -1 \}
\ee
We will denote $ \underbrace{1 \op 1 \op 1\op \cdots \op 1 }_{ n~times} $ by $n_q $. Here we call $n_q$ a q-number of $n$, where $n_q$ reduces to $n$ when $q$ goes to 1.
For real number $x$, we can define the q-number $x_q$ as follows:
\be
x_q =  \qi \{ (2-q)^x -1 \}
\ee
Here we have the following:
\be
0_q = 0 , ~~~ 1_q =1
\ee
Then q-number satisfies the following :
\be
x_q \op y_q = ( x +y )_q
\ee
For this addition, we have the identity $0_q$ obeying
\be
x_q \op 0_q =  x_q
\ee
Letting the inverse of $x_q$ by $(-x)_q $, we have
\be
x_q \op  (-x)_q  =  0_q
\ee
The q-sum satisfies the following property.
\be
(x_1)_q \op (x_2)_q \op \cdots \op (x_n)_q = \frac{ (2-q)^{ \sum_{i=1}^n x_i } -1 }{1-q} = ( \sum_{i=1}^n x_i )_q
\ee

The q-difference is defined in a similar way :
\be
x_q \om  y_q  =  x_q \op  (-y)_q  = (x-y)_q
\ee
The new q-product $\ott$ is defined in [13] as follows:
\be
x \ott y = \frac{ (2-q)^{ \frac{\lx \ly}{\lc^2 }} -1 }{1-q}
\ee
Indeed, $ x \ott y$ reduces to $xy$ when $q$ goes to 1. For this q-product, we have the following:
\be
x_q \ott y_q = (xy)_q
\ee
For q-sum and q-product, the distributive law holds:
\be
x_q \ott (y_q \op z_q ) = (x_q \ott y_q ) \op ( x_q \ott z_q ) = (x(y+z))_q
\ee

The q-product of $n$ variables is defined by
\be
x_1 \ott x_2 \ott \cdots \ott x_n  = \qi \lb (2-q)^{ \frac{ \prod_{i=1}^n \ln (1+(1-q)x_i ) }{ ( \ln (2-q) )^2 } } -1 \rb
\ee
The formula is applied to $n$ q-numbers as follows:
\be
(x_1)_q \ott (x_2)_q \ott \cdots \ott (x_N)_q  = \qi \lb (2-q)^{ \prod_{i=1}^n x_i } -1 \rb = (\prod_{i=1}^n x_i )_q
\ee
When $n$ variables are same, the eq.(21) becomes
\be
x^{\ott^n } = \frac{ (2-q)^{ \left[ \frac{\lx }{\ln(2-q) } \right]^n } -1 }{1-q}
\ee
The eq.(22) is easily proved by mathematical induction. Assume that the eq.(22) holds for $n$. For $n+1$, we have
\be
x^{\ott^{n+1} } = x^{\ott^n } \ott x = \frac{ (2-q)^{ \left[ \frac{\lx }{\ln(2-q) } \right]^{n+1} } -1 }{1-q}
\ee
Thus, for all integers $n$, the eq.(22) holds.

Replacing $x$ with $x_q $ in the eq.(22), we have
\be
x_q^{\ott^n } = (x^n )_q
\ee

Using the eq.(24), we have the following q-binomial theorem.

\be
(x+y)_q^{\ott^n } =( x_q \op y_q )^{\ott^n } = \frac{ (2-q)^{(x+y)^n} -1 }{1-q}
\ee

From the new q-product, we obtain the inverse of $x_q$ , denoted by $\xa $, as follows:
\be
\xa = \qi \lb (2-q)^{x^{-1}} -1 \rb
\ee
Indeed, $\xa$ is an inverse of $x_q $ because
\be
x_q \ott  \xa  = 1
\ee

The q-factorial is defined by
\be
n_q ! = 1_q \ott 2_q \ott \cdots \ott n_q = \frac{(2-q)^{n!} -1 }{1-q}
\ee
The recurrence relation of the q-factorial is then given by
\be
(n+1)_q! = \qi [ (1+(1-q) n_q! )^{n+1} -1 ]
\ee

With a help of the inverse , we can define the q-ratio as follows :
\be
x_q \od y_q = x_q \ott (y^{-1})_q = \frac{ (2-q)^{ \frac{\lx \ln( 1+1-q)(y^{-1})_q ) }{\lc^2 }} -1 }{1-q}
\ee
For the q-ratio, the following holds:
\be
x_q \od  y_q = (xy^{-1})_q
\ee

Form the definition of q-number, we have the following property:
\be
(x+ky)_q = ( 1 + (1-q)(ky)_q ) x_q + ( ky)_q
\ee
or
\be
(x+ky)_q = ( 1 + (1-q)x_q ) ( ky)_q+x_q ,
\ee
where $k$ is an arbitrary real number.

\section{New q-exponential and q-logarithm }

In this section we will investigate the q-exponential and q-logarithm using the q-sum and q-product given in section II. The q-exponential can be expressed in two ways.
\be
e_q(x) = \lim_{n\rightarrow \infty} \lb ( 1 + \frac{x}{n} )^n \rb_q
\ee
or
\be
e_q(x) = \lim_{n\rightarrow \infty} \lb  1 \op  \frac{x}{n} \rb^{\ott n }
\ee
Let us discuss the first definition (34). Then, the q-exponential is given by
\be
e_q(x) = (e^x )_q = \frac{(2-q)^{e^x} -1 }{1-q}
\ee
Indeed, we have
\be
\lim_{q \rightarrow 1 } e_q(x) = e^x
\ee
For the q-product of q-exponentials , we have
\be
e_q(x) \ott e_q (y) = e_q ( x + y )
\ee

As an inverse function of the q-exponential, we can define the q-logarithm as follows:
\be
\la = \ln \lb \frac{\lx }{\ln ( 2-q)} \rb
\ee
Indeed, we have
\be
\lim_{q \rightarrow 1 } \la = \ln x
\ee
The q-logarithm has the following property:
\be
\ln_q ( x \ott y ) = \la + \ln_q y
\ee
\vspace{1cm}

For the second definition (35), the q-exponential has the following form :
\be
e_q(x) = \frac{ (2-q)^{ e^{\frac {(1-q)x}{\ln (2-q)}}} -1}{1-q}
\ee

The proof is easy. From the eq.(35), we have
\bea
e_q(x) &=& \lim_{n\rightarrow \infty} \lb  1 \op  \frac{x}{n} \rb^{\ott n } \cr
&=& \lim_{n\rightarrow \infty} \lb  1 + (2-q) \frac{x}{n}  \rb^{\ott n } \cr
&=& \lim_{n\rightarrow \infty} \qi \left[ (2-q)^{ \lb 1 + \frac{1}{\ln(2-q)} \ln ( 1 + (1-q) \frac{x}{n} ) \rb^n }-1 \right] \cr
&=&  \frac{ (2-q)^{e^{ \frac {(1-q)x}{\ln (2-q)}}} -1}{1-q},
\eea
where we used
\bd
\lim_{n\rightarrow \infty}  \lb 1 + \frac{1}{\ln(2-q)} \ln ( 1 + (1-q) \frac{x}{n} ) \rb^n  = e^{ \frac {(1-q)x}{\ln (2-q)}}
\ed
Indeed, we have
\be
\lim_{q \rightarrow 1 } e_q(x) = e^x
\ee

For the q-product of q-exponentials , we have
\be
e_q(x) \ott e_q (y) = e_q ( x + y )
\ee

As an inverse function of the q-exponential, we can define the q-logarithm as follows:
\be
\la = \frac{\ln(2-q)}{1-q} \ln \lb \frac{\lx }{\ln ( 2-q)} \rb
\ee
Indeed, we have
\be
\lim_{q \rightarrow 1 } \la = \ln x
\ee
The q-logarithm has the following property.
\be
\ln_q ( x \ott y ) = \la + \ln_q y
\ee

\section{q-probability and q-additivity of the q-entropy}
In this section we discuss the q-additivity of the q-entropy which is expressed by using the idea of q-number. Let $W$ be the number of microstates. As the probability of a given microstate among $W$ different ones, we adopt $(p_i)_q $ instead of $p_i $, which is defined by
\be
(p_i)_q = \frac{(2-q)^{p_i} -1}{1-q}
\ee
It is easy to check that $(p_i)_q $ becomes $p_i $ when $q$ goes to 1 . The eq.(49) is a kind of mapping from $p_i$ into $(p_i)_q $. When the condition $\sum_{i=1}^W p_i =1 $ , we have
\be
\bigoplus_{i=1}^W (p_i)_q =1
\ee
where $ \bigoplus_{i=1}^N a_i $ is defined by
\be
\bigoplus_{i=1}^N a_i = a_1 \op a_2 \op  \cdots \op a_N
\ee
Thus, the sum of the probability of each microstate is still unity under replacing an ordinary addition with q-sum and replacing $p_i$ with $(p_i)_q $.

Now let us consider the binomial distribution. At each trial, we may set
\be
p_1 + p_2 =1,
\ee
where $p_1$ is a probability that a certain event occurs at each trial.
If we use a mapping given in the eq.(49), we have
\be
(p_1)_q \op (p_2)_q = 1
\ee
and
\be
[(p_1)_q \op (p_2)_q ]^{\ott n } = 1
\ee
The q-binomial expansion is then as follows:
\be
[(p_1)_q \op (p_2)_q ]^{\ott n } = \bigoplus_{r=1}^n \left[ \frac{1}{1-q} \{ ( 1 + (1-q)p_1^{\ott r } \ott p_2^{\ott(n-r)})^{{}_nC_r } -1 \} \right]
\ee
Let us denote the q-binomial distribution by $B_q(n,p) $ corresponding to the ordinary binomial distribution $B(n,p)$. The probability the the event occurs $r$ times among $n$ trials is given by
\be
(P_r(n))_q =  \frac{1}{1-q} \{ ( 1 + (1-q)p_1^{\ott r } \ott p_2^{\ott(n-r)})^{{}_nC_r } -1 \}
\ee
Then we have
\be
\bigoplus_{r=1}^n (P_r(n))_q = 1
\ee
If we define the q-expectation value of $A$ by
\be
<A>_q = \bigoplus_{r=1}^n (AP_r(n))_q
\ee
Using this, the q-mean is given by
\be
m_q = <r>_q = \frac{ (2-q)^{np_1} -1 }{1-q}
\ee
and q-variance is given by
\be
V_q =  \frac{ (2-q)^{np_1 p_2 } -1 }{1-q}
\ee

Following the above discussion, we can define the q-entropy as follows :
\be
S_q = - k \bigoplus_{i=1}^W ( (p_i)_q \ott (\ln p_i )_q ) = - k ( \sum_{i=1}^W p_i \ln p_i )_q
\ee
Indeed the q-entropy given in the eq.(61) is not additive but q-additive. It obeys the following :
\be
S_q (A+B) = S_q (A) \op S_q (B)
\ee

\section{Conclusion}

In this paper, we constructed two types of new exponential and logarithm related to the non-extensive statistical physics by using the q-sum and q-product which satisfy the distributivity. And we discussed the q-mapping from an ordinary probability to q-probability and investigated the q-binomial distribution. The q-entropy defined through the idea of q-probability was shown to be q-additive.

\def\JMP #1 #2 #3 {J. Math. Phys {\bf#1},\ #2 (#3)}
\def\JP #1 #2 #3 {J. Phys. A {\bf#1},\ #2 (#3)}


\section*{Refernces}

[1] C.Tsallis, J.Stat.Phys.  {\bf 52} (1988) 479.

[2] E.Curado, C.Tsallis, J.Phys. {\bf A 24} (1991) L69.

[3] A.Cho, Science {\bf 297} (2002) 1268.

[4] S. Plastino, Science {\bf 300 } (2003) 250.

[5] L.Nivanen, A.Le Mehaute , Q.Wang, Rep.Math.Phys.{\bf 52} (2003) 437

[6] E. Borges, Physica {\bf A 340} (2004) 95.

[7] S.Abe, A.Rajagopal, Science   {\bf 300 } (2003) 249.

[8] V.Latora, A,Rapisarda, A.Robledo, Science {\bf 300 } (2003) 250.

[9] F. Jackson, Mess.Math.  {\bf 38 }, 57 (1909).

[10] C.Tsallis, Quimica Nova  {\bf 17} (1988) 479.

[11] A.Beardon, An Introduction to hyperbolic geometry, Oxford University Press, New York, (1991).

[12] A.Mehaute, R.Nigmatullin, L.Nivanen, Fleches du temps et geometrie fractale. Hermes, Paris, (1998).

[13] T.Lobao, P.Cardoso, S.Pinho, E.Borges, math-ph/0901.4501 (2009).

\end{document}